\documentclass[11pt]{article}

\usepackage{PRIMEarxiv}

\usepackage[utf8]{inputenc} 
\usepackage[T1]{fontenc}    
\usepackage{hyperref}       
\usepackage{url}            
\usepackage{booktabs}       
\usepackage{amsfonts}       
\usepackage{nicefrac}       
\usepackage{microtype}      
\usepackage{lipsum}
\usepackage{graphicx}
\usepackage{amsmath}
\usepackage{caption}
\usepackage{subcaption}
\usepackage{bigints}
\usepackage{multirow}
\usepackage{algorithm}
\usepackage{algorithmic}

\usepackage{xcolor}
\usepackage[normalem]{ulem}

\usepackage{lineno}

\title{Modeling nonlocal behavior in epidemics via a reaction-diffusion system incorporating population movement along a network
}

\author{
Malú Grave \\
  Dept. of Civil Engineering\\
  COPPE/Federal University of Rio de Janeiro\\
  Fundação Oswaldo Cruz – Fiocruz \\
  \texttt{malugrave@nacad.ufrj.br} \\
   \And
   Alex Viguerie \\
  Department of Mathematics\\
  Gran Sasso Science Institute\\
  \texttt{alexander.viguerie@gssi.it} \\
  \And
    Gabriel F. Barros \\
  Dept. of Civil Engineering\\
  COPPE/Federal University of Rio de Janeiro \\
  \texttt{gabriel.barros@coc.ufrj.br} \\
   \And
 Alessandro Reali \\
  Dipartimento di Ingegneria Civile ed Architettura\\
  Università di Pavia \\
  \texttt{alereali@unipv.it} \\
    \And
 Roberto F. S. Andrade \\
Instituto de Física \\ Universidade Federal da Bahia (UFBA) \\
Center of Data and Knowledge Integration for Health (CIDACS) \\
Instituto Gonçalo Moniz, Fiocruz-Ba \\ 
  \texttt{randrade@ufba.br} \\
  \And
 Alvaro L.G.A. Coutinho \\
  Dept. of Civil Engineering\\
  COPPE/Federal University of Rio de Janeiro \\
  \texttt{alvaro@nacad.ufrj.br} \\
}

\begin{document}
\maketitle

\begin{abstract}
The outbreak of COVID-19, beginning in 2019 and continuing through the time of writing, has led to renewed interest in the mathematical modeling of infectious disease. Recent works have focused on partial differential equation (PDE) models, particularly reaction-diffusion models, able to describe the progression of an epidemic in both space and time. These studies have shown generally promising results in describing and predicting COVID-19 progression. However, people often travel long distances in short periods of time, leading to nonlocal transmission of the disease. Such contagion dynamics are not well-represented by diffusion alone. In contrast, ordinary differential equation (ODE) models may easily account for this behavior by considering disparate regions as nodes in a network, with the edges defining nonlocal transmission. In this work, we attempt to combine these modeling paradigms via the introduction of a network structure within a reaction-diffusion PDE system. This is achieved through the definition of a population-transfer operator, which couples disjoint and potentially distant geographic regions, facilitating nonlocal population movement between them. We provide analytical results demonstrating that this operator does not disrupt the physical consistency or mathematical well-posedness of the system, and verify these results through numerical experiments. We then use this technique to simulate the COVID-19 epidemic in the Brazilian region of Rio de Janeiro, showcasing its ability to capture important nonlocal behaviors, while maintaining the advantages of a reaction-diffusion model for describing local dynamics.
\end{abstract}


\section{Introduction}
\label{sec:introduction}
The outbreak of COVID-19, which started in 2019 and is still continuing, has caused unprecedented disruption in terms of both human cost and economic damage. In order to better understand the dynamics of the disease spread, with the hope of ultimately improving policy and public health outcomes, there has been an explosion in the study of mathematical modeling of infectious disease. These models have taken many forms, and a comprehensive survey of the literature is beyond the scope of the current work. However, we note that many different approaches have been used to model the epidemic, including machine-learning and data-driven approaches \cite{zhang2021integrated, viguerie2022coupled, jha2020bayesian, calvetti2020metapopulation, bhouri2021covid}, models using a classical compartmental approach, together with parameter estimation techniques \cite{gatto2020spread, parolini2021suihter}, delay differential equations \cite{guglielmi2021delay}, partial differential equations \cite{bertaglia2021hyperbolic, bertrand2021least, viguerie2020diffusion, zanella2021data}, network-based methods \cite{miranda2021scaling, kuhl2021computational,linka2020safe,bertaglia2021hyperbolic}, as well as agent-based \cite{zohdi2020agent, kerr2021covasim} and multiscale models \cite{bellomo2020}. We note that the cited works represent just a small sample of the total literature, and further that the various approaches discussed are not mutually exclusive. For a recent survey, see \cite{kuhl2021computational}.

\par The basis for most mathematical models of infectious disease is rooted in the compartmental theory of  Kermack and McKendrick \cite{kermack1927contribution} (see \cite{breda2012formulation} for a more modern treatment). This approach divides a given population into different states corresponding to disease status, with the underlying equations governing the evolution of the population composition in terms of disease-state. This mechanism forms the backbone behind the classic S-I-R (\textit{susceptible-infected-removed}) model and its variants: for instance, the S-I-S (\textit{susceptible-infected-susceptible}) and S-E-I-R (\textit{susceptible-exposed-infected-removed}), among others. While the initial model presented by Kermack and McKendrick is quite general, and presented in the form of an integral renewal equation, under certain assumptions, namely a mass-action transmission mechanism and exponentially-distributed sojourn times, we obtain the familiar systems of ordinary differential equations (ODE) widely used today. These systems should be recognized, however, as special cases within a far more general framework \cite{kermack1927contribution, breda2012formulation}. We note that many models, including agent-based and evolving network models, even if not expressed as integral or differential equations, still generally utilize some sort of compartmental structure \cite{sonza2021progression, zohdi2020agent}.
\par Such compartment models have historically been applied with great success in the modeling of infectious disease, and offer a clear mechanistic description of epidemic progression. However, in their most widely-utilized ODE formulations, 
relevant factors may not be taken into account. Among the most important such factors, and the subject of this work, is the introduction of \textit{spatial information} into the epidemic model. Many approaches have been proposed to describe epidemic evolution in space as well as time. A common approach is to further stratify the compartmental structure by geographic location, such that different compartments also correspond to different geographic areas, often delimitated by political jurisdiction. For COVID-19, this approach has been utilized in e.g. \cite{miranda2021scaling,gatto2020spread, bhouri2021covid, linka2020safe} and offers the advantage of computational efficiency and relative ease of implementation. Additionally, the spatial evolution is not inherently local, and nonlocal processes (ie, transmission between individuals in distant regions) can be easily incorporated. However, a continuous description of spatial dynamics with this approach is impossible, and as the spatial resolution increases, the number of necessary compartments can quickly become intractable. 
\par In contrast, \textit{partial differential equations} (PDE) incorporating the familiar compartmental structure have been offered as an alternative. Most commonly, models of this type incorporate evolution in space via a reaction-diffusion equation \cite{MurrayII, keller2013numerical}. Works incorporating PDE can be seen in   \cite{viguerie2020diffusion, viguerie2021simulating, jha2020bayesian, bertaglia2021hyperbolic, bertrand2021least, grave2020new, grave2021assessing}. Most of these used a reaction-diffusion process, with \cite{bertaglia2021hyperbolic} introducing a multiscale approach utilizing convection for nonlocal processes and diffusion for local processes. In \cite{ramaswamy2021comprehensive} convection is also employed to account for population mobility.
\par PDE models offer the large advantage of a true, spatially continuous description; however, they too have downsides. From the practical point of view, they are generally more difficult to implement and require greater computational resources than ODE models. From the modeling point of view, the ability of a linear diffusion model to accurately describe the spatial evolution of an epidemic in human populations, given the nature of human movement, is questionable. While many of the COVID-19 related applications of these models mitigate this problem somewhat through, for instance, nonlinear diffusion and depensation effects \cite{viguerie2020diffusion, viguerie2021simulating, jha2020bayesian, bertrand2021least, grave2020new, grave2021assessing}, the model still behaves locally. Although \cite{bertaglia2021hyperbolic} does make an excellent attempt at a nonlocal description, ultimately, despite the presence of multiple scales of spatial evolution and convection, the underlying process is still local, as contagion is still spread along with the intermediate points between origin and destination. It is well-known that bilaplacian \cite{MurrayII} or fractional diffusion \cite{d2013fractional} operators can describe nonlocal dynamics. Some works have indeed applied such techniques to COVID-19 (see \cite{higazy2020novel, abadias2020fractional}); however, the complexity of these models in terms of computational effort and stability, as well as their still-developing theory, makes their widespread adoption, at least at the current time, difficult. Further, the nonlocal movement in human populations is far from random; indeed, it often follows a predictable pattern \cite{Alla2020}.
\par In contrast to these approaches, we propose herein a combination of the reaction-diffusion model first proposed in \cite{viguerie2020diffusion} with network-based methods, akin to those proposed in \cite{miranda2021scaling, kuhl2021computational,linka2020safe,bertaglia2021hyperbolic}. Similarly to the modeling paradigm introduced in \cite{bertaglia2021hyperbolic}, we employ a PDE model and postulate that diffusion occurs at the local (i.e. \textit{within-city}) level. However, rather than defining nonlocal dynamics via convection, we instead model such dynamics as sources and sinks within a network. We embed a spatial network structure within the underlying PDE system, and define a network via transport operators between different areas in the spatial domain (i.e., inter-city mobility). These operators then allow portions of the population to, for instance, move from one city to another without diffusing the disease along the way, as is typically the case with vehicular transport. We believe that such a method offers a reasonable, easily implemented, stable approach that maintains the advantage of a PDE-based model for local dynamics while also incorporating important nonlocal effects.

\par We outline the article as follows: we first introduce and provide a brief explanation of the underlying equations. We then discuss the idea behind the population transfer network and  its mathematical formulation, as well as important details regarding its numerical implementation. We then provide numerical examples on an idealized problem, as well as a realistic problem in the Brazilian state of Rio de Janeiro, to provide evidence of the model's ability to describe the relevant local and nonlocal dynamics found in human epidemics.

\section{Governing equations}\label{sec:methods}

The governing equations are based on the a spatio-temporal SEIRD model, presented in \cite{viguerie2020diffusion, viguerie2021simulating, grave2020adaptive}. We consider that only the susceptible and exposed compartments are able to move along the network. The system of equations becomes:

\begin{equation}\begin{split}
&\frac{\partial s}{\partial t} + \frac{\beta_i}{n} si + \frac{\beta_e}{n}  se 
- \nabla \cdot (n\nu_s\nabla s) +  S_s s = 0
\end{split}
\label{covid_s}
\end{equation}
\begin{equation}
\begin{split}
&\frac{\partial e}{\partial t} - \frac{\beta_i}{n} si - \frac{\beta_e}{n}se + (\alpha  + \gamma_e) e 
- \nabla \cdot (n\nu_e \nabla e) +  S_e e = 0
\end{split}
\label{covid_e}
\end{equation}
\begin{equation}
\frac{\partial i}{\partial t} - \alpha e + (\gamma_i  + \delta) i - \nabla \cdot (n\nu_i \nabla i) = 0
\label{covid_i}
\end{equation}
\begin{equation}
\frac{\partial r}{\partial t}  -\gamma_e e- \gamma_i i  -\nabla \cdot (n \nu_r \nabla r) = 0
\label{covid_r}
\end{equation}
\begin{equation}
\frac{\partial d}{\partial t} - \delta i = 0,
\label{covid_d}
\end{equation}

\noindent where $s(\mathbf{x}, t)$, $e(\mathbf{x}, t)$, $i(\mathbf{x}, t)$, $r(\mathbf{x}, t)$, and $d(\mathbf{x}, t)$ denote the densities of the \textit{susceptible}, \textit{exposed}, \textit{infected}, \textit{recovered}, and \textit{deceased} populations, respectively. $\beta_i$ and $\beta_e$ denote the \textit{contact rates} between symptomatic and susceptible individuals and asymptomatic and susceptible individuals, respectively (units days$^{-1}$),
$\alpha$ denotes the incubation period (units days$^{-1}$), $\gamma_e$ corresponds to the asymptomatic recovery rate (units days$^{-1}$), $\gamma_i$ the symptomatic recovery rate (units days$^{-1}$), $\delta$ represents the mortality rate (units days$^{-1}$), and $\nu_s$, $\nu_e$, $\nu_i$, $\nu_r$ are the diffusion parameters of the different population groups as denoted by the sub-scripted letters (units km$^2$  persons$^{-1}$  days$^{-1}$). Finally, $S_s$ and $S_e$ are \textit{population transfer operators}, defining the movement of populations within the population transfer network as sources and sinks in the reaction-diffusion system. The precise definition of these operators will be discussed in detail in the following section. The system of equations \eqref{covid_s}-\eqref{covid_d} is supplemented with proper boundary and initial conditions.

The movement occurs between paired regions, ie, nodes that share an edge in the underlying network. Movement away from one region results in movement into the other region. In this sense, in a given time, when persons move away from a region, the population transfer operator acts as a \textit{sink} term; likewise, when persons move into a region, the population transfer operator acts as a \textit{source} term.

\section{Defining the transfer operator}
\label{sec:analysis}
In this section we define the population transfer operators $S$ within a network. For simplicity, we consider a single compartment and a network consisting of two nodes; extending the definitions to larger networks and other compartments follows immediately.
\subsection{Model problem}
We consider the equation \eqref{covid_s},
\begin{equation}
    \frac{\partial s}{\partial t} + \frac{\beta_i}{n} si + \frac{\beta_e}{n}  se 
- \nabla \cdot (n\nu_s\nabla s) +  S_s s = 0,
\end{equation}
defined in a domain $\Omega$, with two distinct subsets $\Omega_1 \subset \Omega$, $\Omega_2 \subset \Omega$ such that $\Omega_1 \cap \Omega_2 = \emptyset.$ 
\subsection{Transfer network definition}
We define the directed, weighted, and without self-loops \textit{transfer network} as $N=G(\mathbf{\Omega},\mathbf{\eta})$, where $\mathbf{\Omega}$ and $\mathbf{\eta}$ represent, respectively, the set of nodes $\Omega_i$ and edges $\eta_i^j$ between nodes $i$ and $j$. For the time being, let $N$ consist of only two nodes, i.e. $i=1,2$. We also assume that $N$ is a \textit{time-varying graph} (TVG) \cite{nicosia2012}, where the edges $\eta_{ij}$ depend on time according to the following assumptions (note this dependence on time is assumed if not explicitly denoted):
\begin{enumerate}
\item Every $T_1^2$ units of time, a percentage $p_1^2$ of the population in $\Omega_1$ move to $\Omega_2$;
\item Every $T_2^1$ units of time, a percentage $p_2^1$ of the population in $\Omega_2$ move to $\Omega_1$.
\end{enumerate}
This is equivalent to stating that the probability that an individual moves from $\Omega_1$ to $\Omega_2$ (or vice versa) over $T_1^2$ time units is $p_1^2$.
\par For example, if one assumes that this probability is distributed exponentially with parameter $\eta_1^2$, we find that:
\begin{align}\begin{split}\label{movementProb}
\int_0^{T_1^2} \eta_1^2 e^{-\eta_1^2 t} dt &= p_1^2 \\
1-e^{-\eta_1^2 T_1^2} &= p_1^2,
\end{split}\end{align}
which after some straightforward algebra gives:
\begin{align}\label{rate1to2}
    \eta_1^2 &= -\frac{\ln(1-p_1^2)}{T_1^2}.
\end{align}

\subsection{Transfer operator definition}
We now define the \textit{population transfer operator along the network} $S_s(N)$. The dependence of $S_s(N)$ on the network $N$ is assumed and not denoted explicitly in hereafter. Such an operator is not unique, and in this section we outline the properties that an operator of this type must satisfy. We split the operator into two components, a \textit{donor operator} $D_s$ which transfers a population from $\Omega_i$ to $\Omega_j$ at a rate $\eta_i^j$ and the corresponding \textit{receiver operator} $R_s$ which receives a population from $\Omega_i$ into $\Omega_j$ at a rate $\eta_i^j$. $S_s$ is then given as the sum of these components: $S_s = D_s +R_s$. $S_s \lbrack\,\cdot\, \rbrack$ (and its individual components) is a linear operator in $\lbrack\,\cdot\, \rbrack$, although it may have additional dependencies on $\mathbf{x}$, any of the compartments, and/or $t$ in general, and need not be linear in these variables. \par $S_s$ must satisfy the following conditions:
\begin{enumerate}
    \item \textit{Non-negativity of the donor operator}: $(I+D_s )s \geq 0 $ for all $(\mathbf{x},t)$;
    \item \textit{Population (mass)-conservation}: $\int_{\Omega} S_s  s \,d\Omega = 0 $ for all $t$.
\end{enumerate}
Both conditions are physically motivated; the first ensures that the quantity of persons transferred away from given node $\Omega_i$ must not exceed the total persons present in $\Omega_i$. Additionally, 
this condition is necessary to guarantee the coercivity (and thereby well-posedness) of the variational problem corresponding to \eqref{covid_s}-\eqref{covid_d} (see e.g. \cite{auricchio2022well}).
\par The population conservation condition is necessary to guarantee that the transfer operator only models movement among the different regions, and does not change the overall quantity of persons (or mass) in the system.

We now discuss the donor and receiver operators separately.

\subsubsection{Donor operator}
In the movement of populations from $\Omega_1$ to $\Omega_2$ (and vice versa) at the rate $\eta_1^2$ (and $\eta_2^1$), the \textit{donor operator} $D_s$ defines the removal of persons from the origin node.
\par We propose the following, simple definition of $D_s$:
\begin{align}\label{donorFun}
    D_s(\mathbf{x})\lbrack\, \cdot\, \rbrack = -\eta_1^2 \chi_{\Omega_1}(\mathbf{x})\lbrack\, \cdot\, \rbrack   - \eta_2^1 \chi_{\Omega_2}(\mathbf{x}) \lbrack\, \cdot\, \rbrack,
\end{align}
where $\chi_{\Omega_i}(\mathbf{x})$ are characteristic functions:
\begin{align}\label{chiDef}
\chi_{\Omega_i}(\mathbf{x}) &= \begin{cases} 1 \,\,\, \text{if } \mathbf{x} \in \Omega_i, \\
0 \,\,\, \text{else.}
\end{cases}
\end{align}
Assuming the $\eta_i^j < 1$, it follows immediately from $\Omega_1 \cap \Omega_2 = \emptyset$ that $(I+D_s)s>0$, ensuring that the non-negativity condition is satisfied.

\subsubsection{Receiver operator}
The \textit{receiver operator} $R_s$ receives the populations taken from the donor region by the donor operator and distributes them in the receiver region, and has the following form:
\begin{align}\label{receiverFun}
R_s\lbrack\,\cdot\,\rbrack &= \eta_1^2 \, \chi_{\Omega_2}(\mathbf{x}) K_{\Omega_2}(\mathbf{x},t)  \int_{\Omega_1} \lbrack\,\cdot\,\rbrack d\Omega_1 + \eta_2^1 \,\chi_{\Omega_1}(\mathbf{x},t) K_{\Omega_1} \int_{\Omega_2} \lbrack\,\cdot\,\rbrack d\Omega_2.   
\end{align}
The \textit{kernel functions} $K_{\Omega_i}(\mathbf{x},t)$ are defined such that:
\begin{align}\label{kernelPos}
    K_{\Omega_i}(\mathbf{x},t) \geq 0,\, \forall\, \mathbf{x} \in \Omega_i,\, \forall t,
\end{align}
and
\begin{align}\label{kernelOne}
 \int_{\Omega_i} K_{\Omega_i}(\mathbf{x},t) \,d\Omega_i &= 1 \,\,\,\forall t.
\end{align}
\textbf{Theorem. } \textit{The transfer function defined by \eqref{donorFun}, \eqref{receiverFun} satisfies the mass-conservation property.}
\newline \textbf{Proof.}
One may verify this with straightforward computation:
\begin{align}\begin{split}\label{nonNegProof}
\int_{\Omega} S_s s \, d\Omega &= \int_{\Omega} D_s s\,  d\Omega + \int_{\Omega} R_s s \,d\Omega \\
&= -\int_{\Omega}\eta_1^2 \, \chi_{\Omega_1}(\mathbf{x}) s d\Omega -  \int_{\Omega}\eta_2^1 \,  \chi_{\Omega_2}(\mathbf{x}) s \, d\Omega\\ 
&\quad+  \int_{\Omega} \left(\eta_1^2\, \chi_{\Omega_2}(\mathbf{x}) K_{\Omega_2}(\mathbf{x},t) \int_{\Omega_1} s \, d\Omega_1 \right) d\Omega  + \int_{\Omega} \left(\eta_2^1 \, \chi_{\Omega_1}(\mathbf{x}) K_{\Omega_1}(\mathbf{x},t) \int_{\Omega_2} s \,d\Omega_2\right)d\Omega \\
&= -\eta_1^2 \int_{\Omega_1} s\, d\Omega - \eta_2^1 \int_{\Omega_2} s\,d\Omega \\
&\quad+ \eta_1^2\int_{\Omega_1} s\, d\Omega_1 \left(\int_{\Omega} \chi_{\Omega_2}(\mathbf{x})K_{\Omega_2}(\mathbf{x},t) \,d\Omega \right) + \eta_2^1\int_{\Omega_2} s\, d\Omega_2 \left(\int_{\Omega} \chi_{\Omega_1}(\mathbf{x})K_{\Omega_1}(\mathbf{x},t)\,d\Omega \right) \\
&=-\eta_1^2 \int_{\Omega_1} s\, d\Omega - \eta_2^1 \int_{\Omega_2} s\,d\Omega \\
&\quad+ \eta_1^2\int_{\Omega_1} s\, d\Omega_1 \left(\int_{\Omega_2} K_{\Omega_2}(\mathbf{x},t) \,d\Omega \right) + \eta_2^1\int_{\Omega_2} s\, d\Omega_2 \left(\int_{\Omega_1}K_{\Omega_1}(\mathbf{x},t)\,d\Omega \right) \\
&= 0.
\end{split}\end{align}
\subsubsection{Some examples of transfer operators}
Transfer operators of the type introduced above are non-unique, and any operator satisfying the necessary properties may be applied in practice. Besides the clearly problem-dependent $\eta_i^j$, the definition of the kernel functions $K_{\Omega_i}(\mathbf{x},t)$, which corresponds to the distribution of the transferred population within the receiver region, is the most important component in constructing $S_s$.
\par As shown in the preceding sections, as long as the conditions \eqref{kernelPos}-\eqref{kernelOne} are satisfied, then one may define an acceptable transfer function. However, these conditions are quite non-restrictive, and many such functions will satisfy them. Defining $K_{\Omega_i}$ is therefore far from straightforward, and problem-specific considerations, computational constraints, and available data are likely to inform such choices in practice. Below we list two possible, obvious definitions.
\par \textbf{Uniform distribution. } A possible definition is given by:
\begin{align}\label{transferUniform}
    K_{\Omega_i}(\mathbf{x},t) &= \frac{1}{m(\Omega_i)}.
\end{align}
This defines the simplest possible transfer function, which distributes the population uniformly among the receiver domain. It is computationally simple and, therefore, the approach applied in the simulations of this work. However, it is questionable how well this definition captures reality in most instances.
\par \textbf{Receiver-proportional distribution.} Another possible, natural choice is:
\begin{align}\label{transferReceiver}
 K_{\Omega_i}(\mathbf{x},t) &= \frac{1}{\int_{\Omega_i} s\,d\Omega_i} s,
\end{align}
which distributes the population in proportion to the existing population density in $\Omega_i$. This is likely more realistic than the uniform distribution; however, it requires the additional evaluation of an integral, must be updated in time, and introduces a nonlinearity into the problem. 
\par Depending on how one chooses to discretize the term \eqref{transferReceiver} in time, it may become necessary to apply some nonlinear iteration scheme at each time step. We note, however, that such schemes are generally necessary when solving \eqref{covid_s}-\eqref{covid_d} anyway, and this may not necessarily represent a large additional computing cost.

\section{Results}
\label{sec:results}

\subsection{Simple test problem}
\begin{figure}[ht!]
    \centering
    \includegraphics[width=0.7\textwidth]{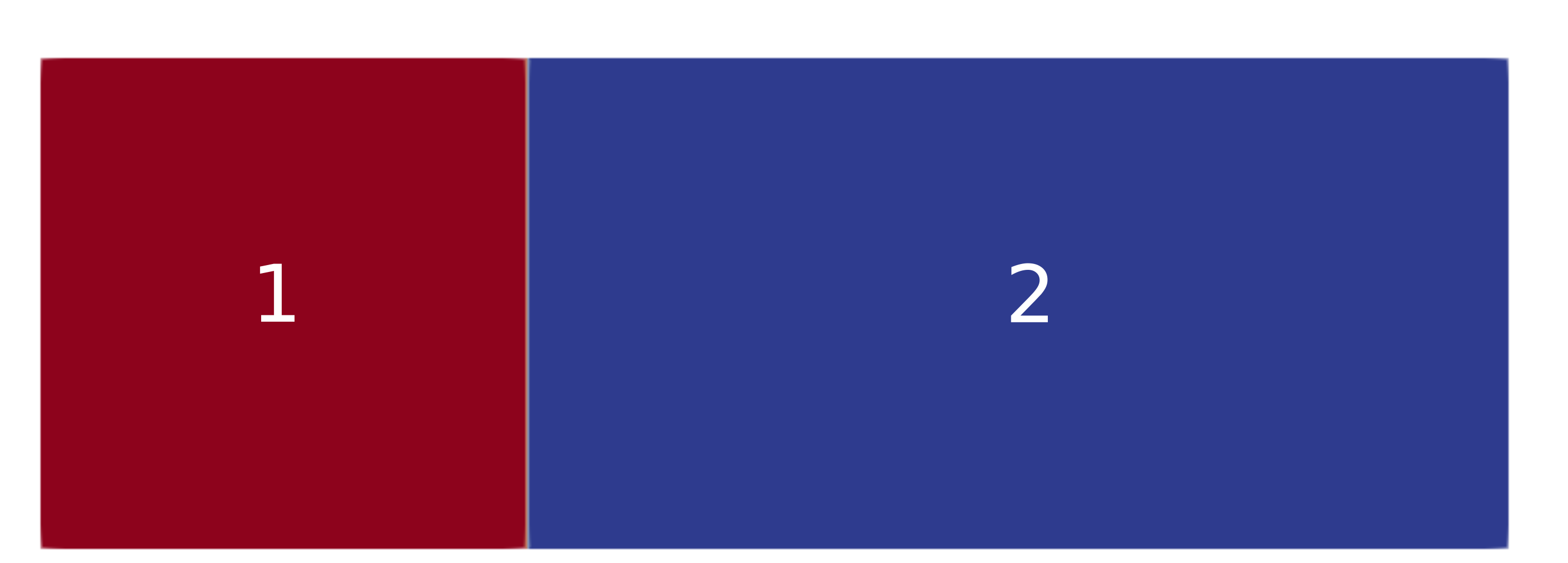}
    \caption{Simple test problem schematic}
    \label{fig:toy}
\end{figure}

In order to check the implementation of the source and sinks algorithm, we use a rectangular domain that is divided into two regions. The dimensions of each region are $1\times 1$ for Region 1 and $2\times 1$ for Region 2. We start with 1000 people in each region, i.e., Region 1 has 1000 $people/km^2$ and Region 2 has 500 $people/km^2$ (Figure \ref{fig:toy}).

The objective is to seasonally send people from Region 1 to Region 2 and people from Region 2 to Region 1. We define a weekly pattern in which people leave Region 1 on Fridays and go back home on Sundays. For instance, assuming day 1 is Monday, people leave Region 1 to Region 2, and on day 7 people leave Region 2 to Region 1. The movement is defined to be 20\% of the population in each region and assumed to be distributed evenly throughout the day. Hence:

\begin{equation}
    \eta_{1}^2 = \begin{cases} 
    0.2/\text{day} \text{ if Friday,} \\
    0/\text{day} \text{ otherwise}
        \end{cases}, \qquad     \eta_{2}^1 = \begin{cases} 
    0.2/\text{day} \text{ if Sunday,} \\
    0/\text{day} \text{ otherwise.}
        \end{cases}
\end{equation}
We define $K_{\Omega_{1,2}}$ as in \eqref{transferUniform}.

\begin{figure}[ht!]
    \centering
    \includegraphics[width=0.7\textwidth]{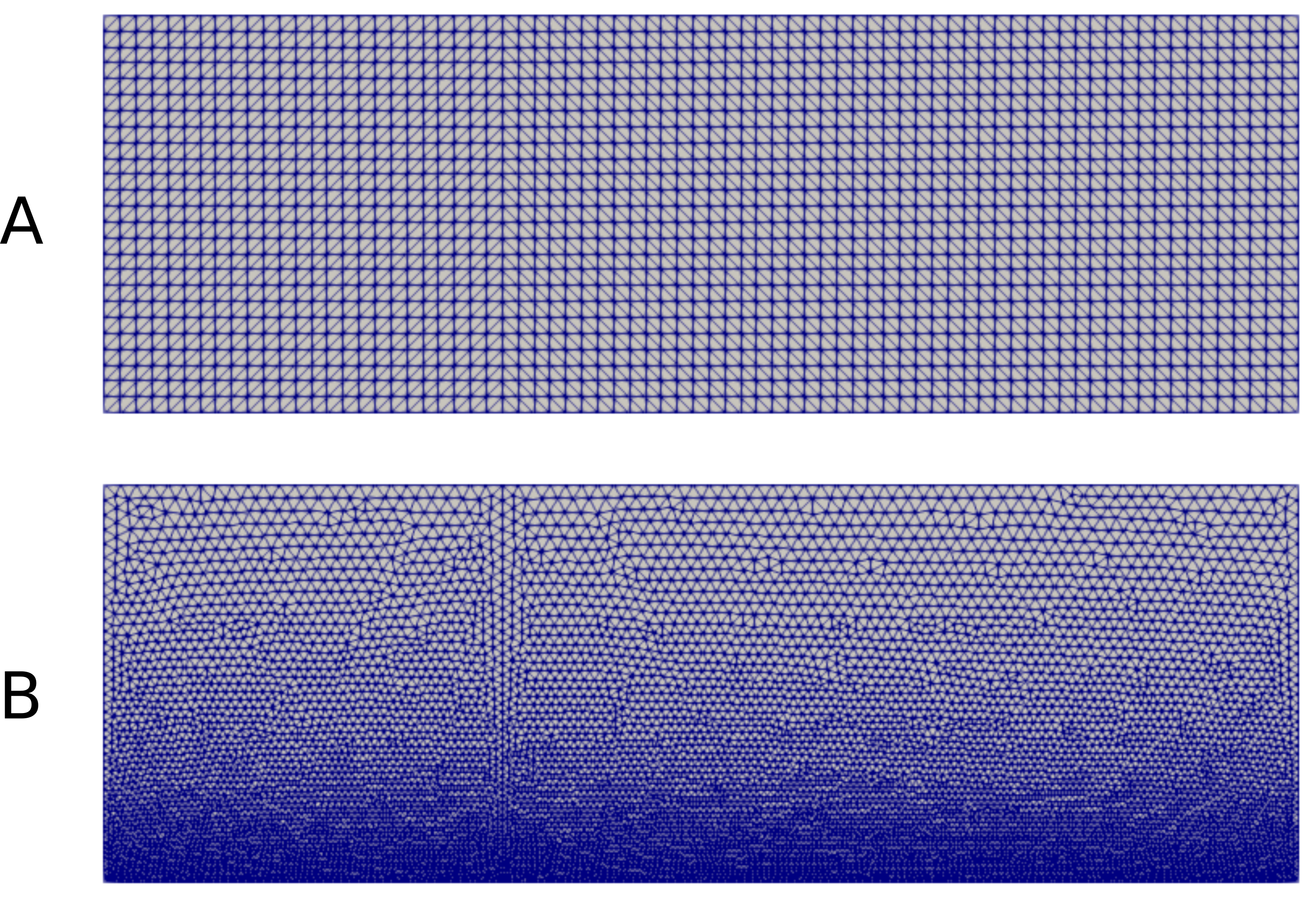}
    \caption{Simple test problem meshes. A) 25$\times$50 elements mesh. B) unstructured mesh.}
    \label{fig:grids}
\end{figure}

We use different meshes to study population (mass) conservation. Three meshes are structured with different element mesh sizes (25$\times$50, 50$\times$100, and 100$\times$200). The fourth mesh is unstructured with 18,231 elements of sizes varying between the bigger and smaller elements of the fixed meshes. All discretizations with linear triangles. In Figure \ref{fig:grids} we show the coarsest structured and the unstructured meshes.

As we are not considering diffusion in this example, the only change in population distribution is due to $S_s$. Therefore, it is possible to determine the analytical solution for this problem.

\begin{figure}[ht!]
    \centering
    \includegraphics[width=\textwidth]{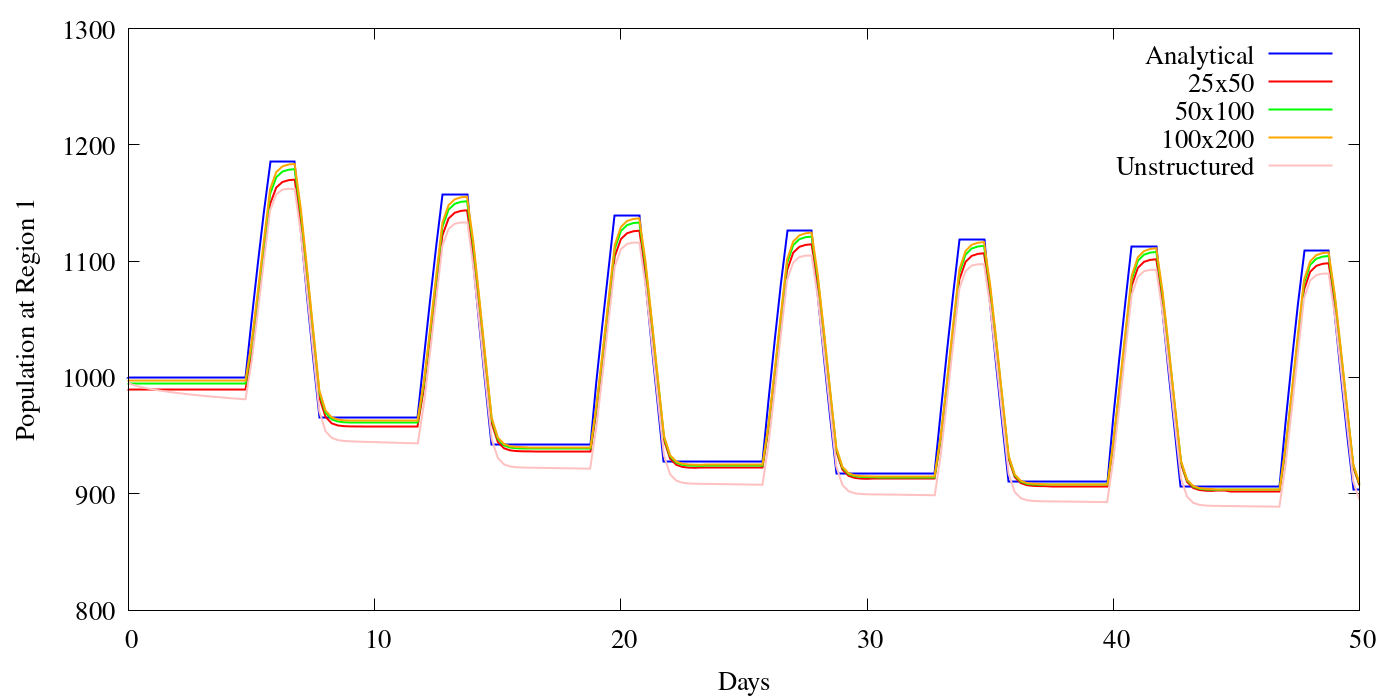}
    \caption{Population at Region 1}
    \label{fig:region1}
\end{figure}

\begin{figure}[ht!]
    \centering
    \includegraphics[width=\textwidth]{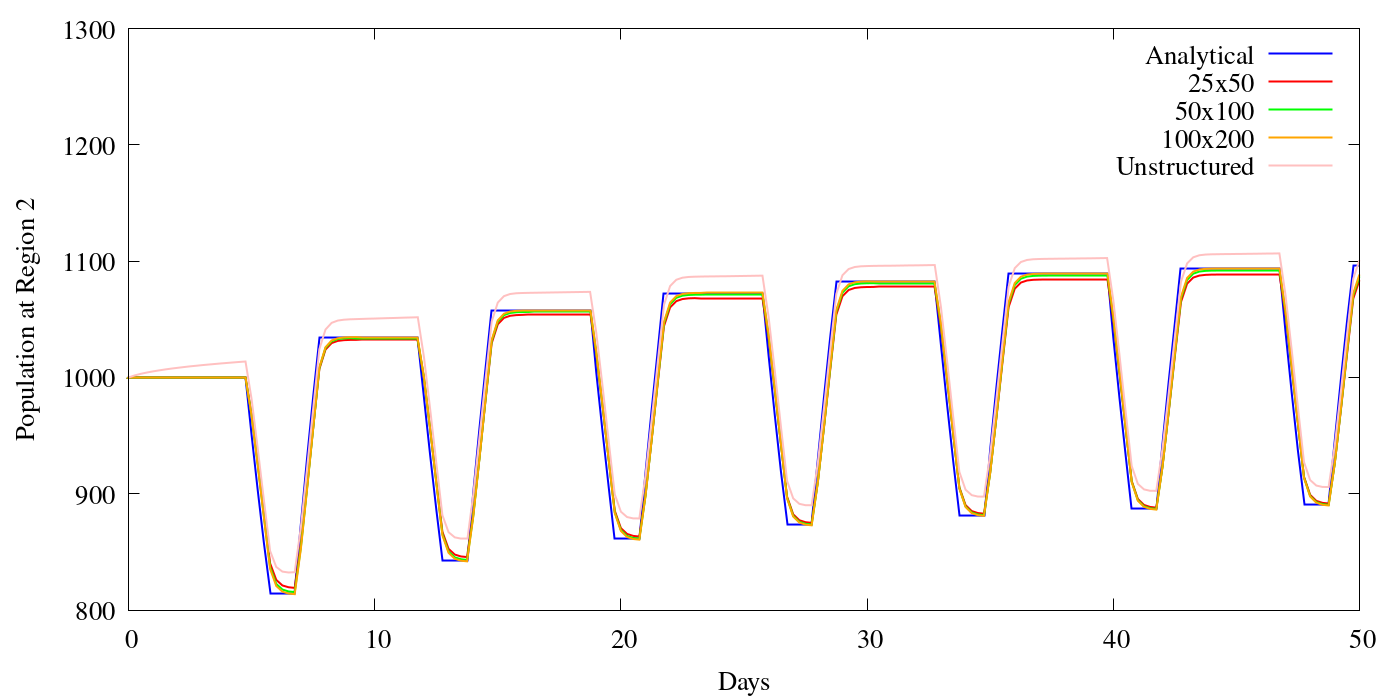}
    \caption{Population at Region 2}
    \label{fig:region2}
\end{figure}

\begin{figure}[ht!]
    \centering
    \includegraphics[width=\textwidth]{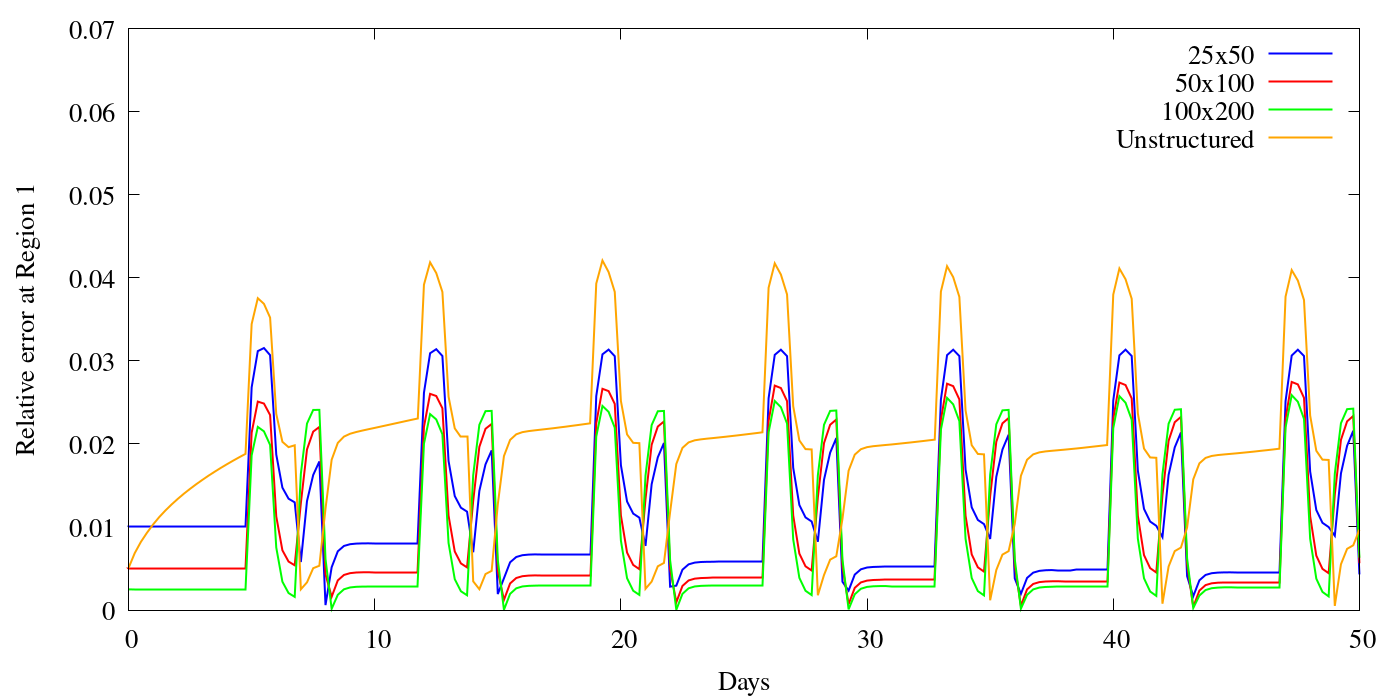}
    \caption{Population relative error at Region 1}
    \label{fig:region1error}
\end{figure}

\begin{figure}[ht!]
    \centering
    \includegraphics[width=\textwidth]{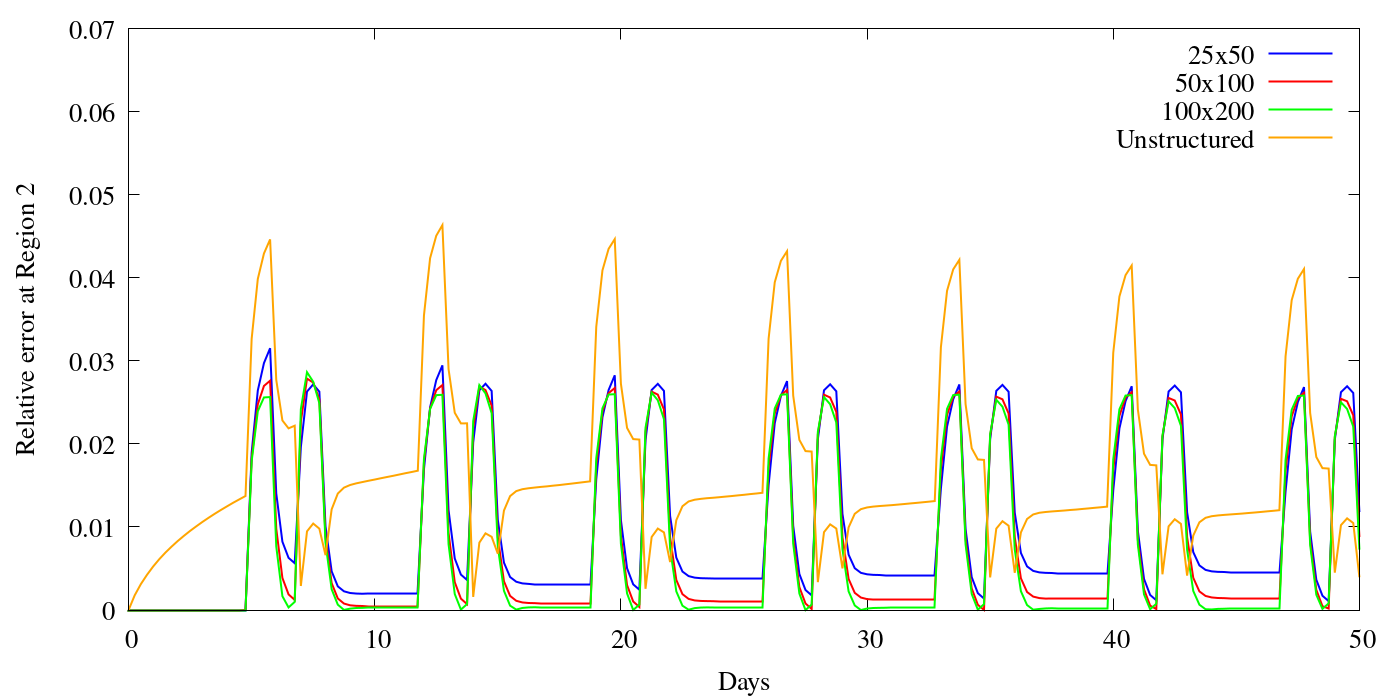}
    \caption{Population relative error at Region 2}
    \label{fig:region2error}
\end{figure}

\begin{figure}[ht!]
    \centering
    \includegraphics[width=\textwidth]{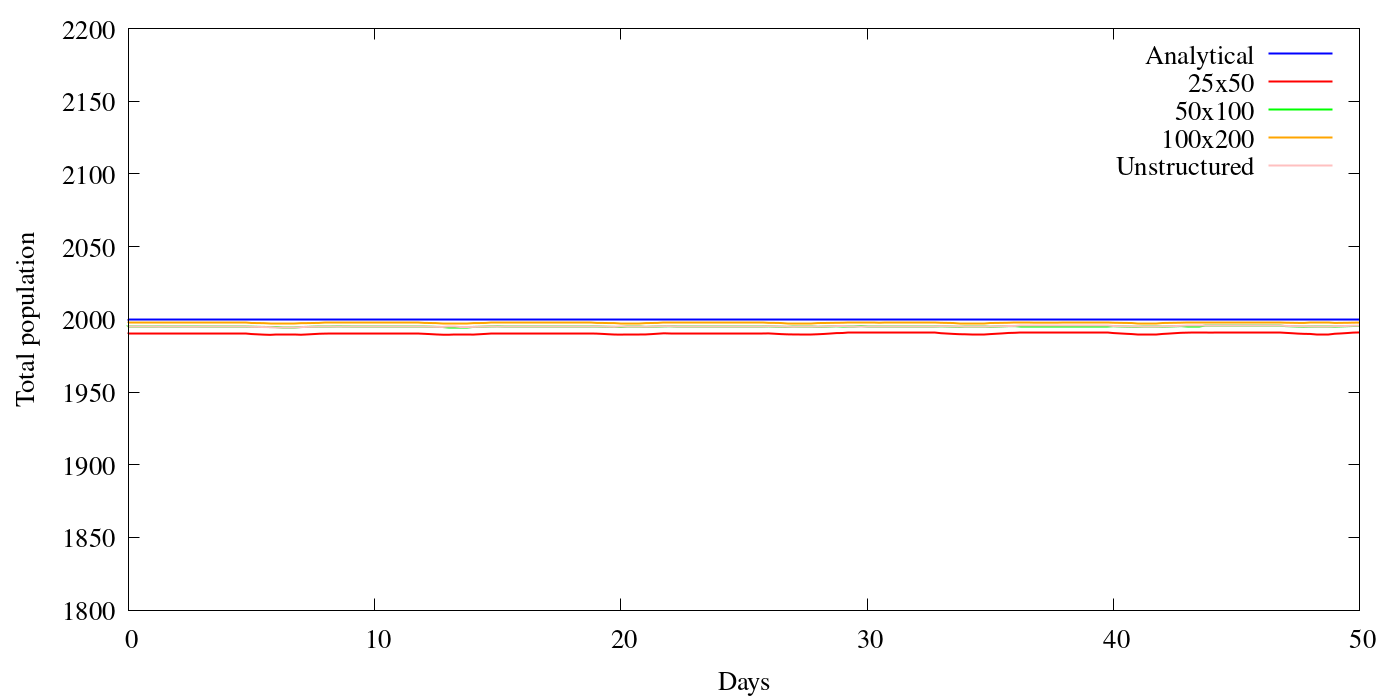}
    \caption{Total population}
    \label{fig:mass}
\end{figure}

\begin{figure}[ht!]
    \centering
    \includegraphics[width=\textwidth]{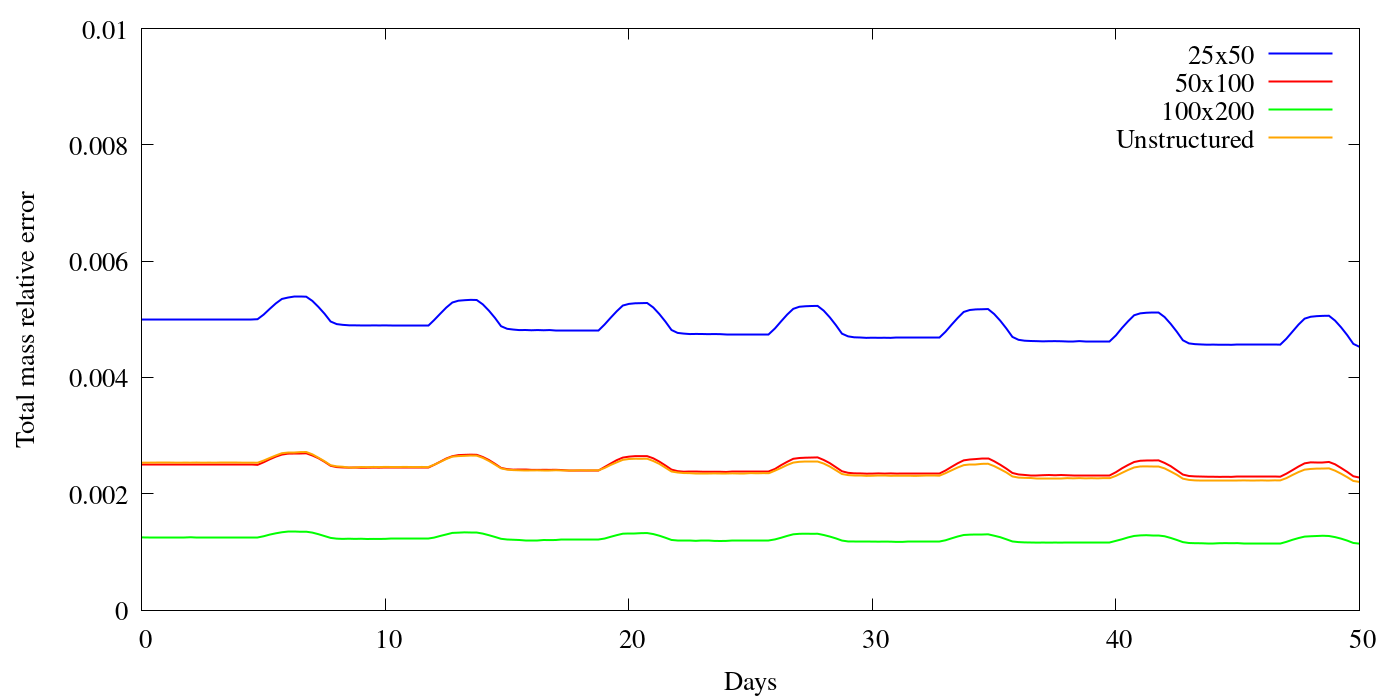}
    \caption{Total population relative error for each grid}
    \label{fig:masserror}
\end{figure}

In Figures  \ref{fig:region1} and \ref{fig:region2} we plot the total populations in each time-step for Region 1 and 2, considering the four meshes. Figures \ref{fig:region1error} and  \ref{fig:region2error} show the relative error between the analytically-computed number of people during each time-step for Region 1 and 2, for each of the four meshes. Finally, Figure  \ref{fig:mass} shows the total population for each mesh and Figure  \ref{fig:masserror} shows the total population (mass) relative error between the simulations and analytical solution.

We briefly note that, as we keep all of the percentages fixed at 20\%, we observe population drift over time, with greater numbers of people being moved each time. This occurs due to the fact that, after sending 20\% of the population in Region 1, when we send 20\% of the population in Region 2 back to Region 1, this population now includes the original persons transferred. Noting that 20\% of 1000 is 200, we may avoid this problem by defining the transfer rates instead as:

\begin{equation}\label{preciseEta}
   \eta_{1}^{2} = \begin{cases} 
    \frac{200}{\int_{\Omega_{1}} s d\Omega_{1} } \cdot \text{ day}^{-1} \text{ if Friday,} \\
    0/\text{day} \text{ otherwise}
        \end{cases}, \qquad     \eta_{2}^1 = \begin{cases} 
   \frac{200}{\int_{\Omega_{2}} s d\Omega_{2}} \cdot \text{ day}^{-1}\text{ if Sunday,} \\
    0/\text{day} \text{ otherwise.}
        \end{cases}
\end{equation}
While defining $\eta$ in such a manner requires integral evaluations, we note that these are already computed in order to define \eqref{receiverFun}. 

Depending on the mesh refinement, we start the simulation with a different amount of total population, close to 2000. This happens because of the way that the initial conditions are set. The more refined the mesh, less difference between the real value and the one provided by the simulation. As we increase the refinement, we decrease the errors of conservation of the total population during the simulation. However, when we look to the populations at each region, we see that the unstructured mesh returns bigger errors than the structured ones.

\subsection{Rio de Janeiro - Brazil}

\begin{figure}[ht!]
    \centering
    \includegraphics[width=0.7\textwidth]{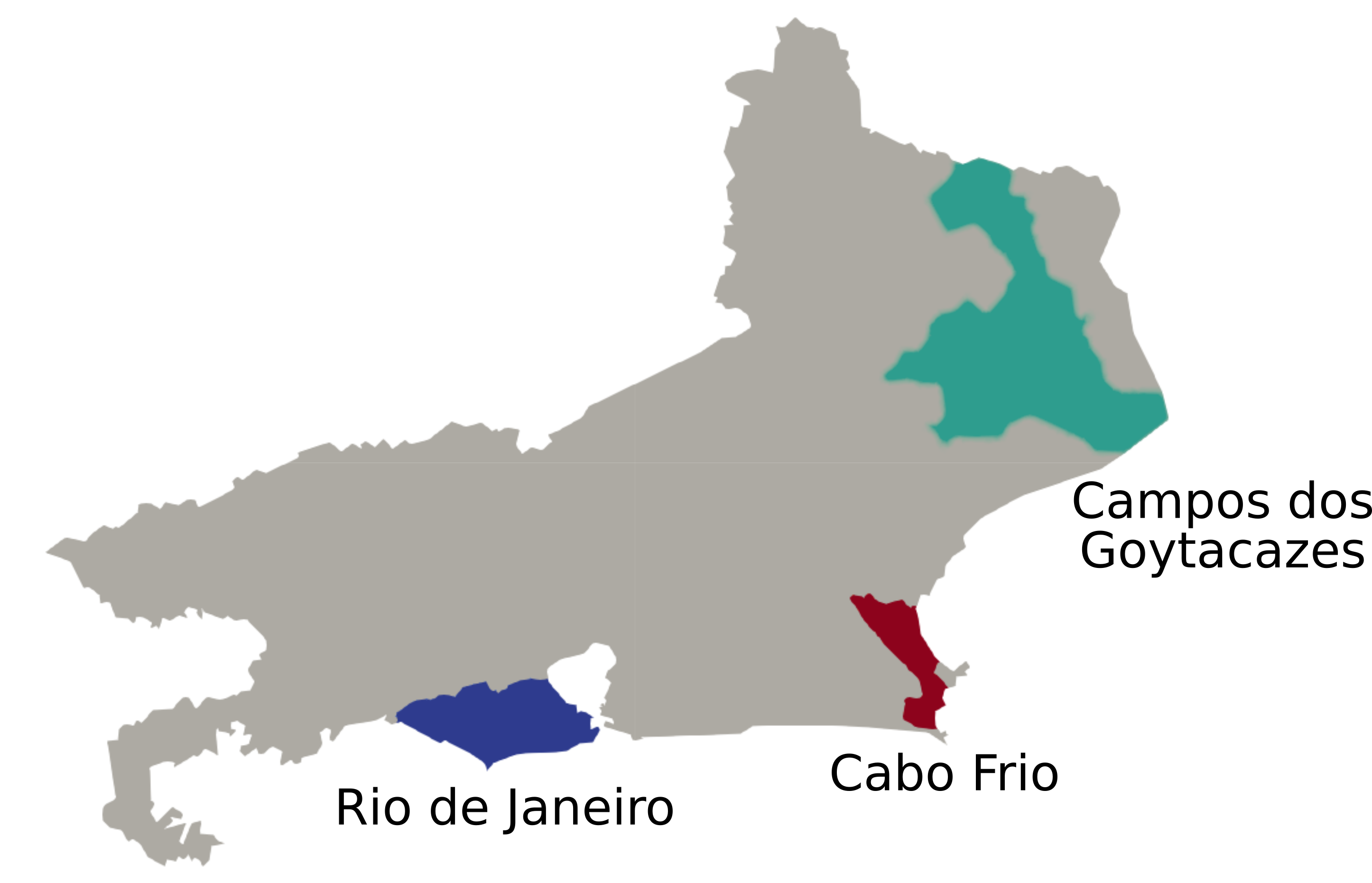}
    \caption{Sketch of the Rio de Janeiro state, Brazil, showing the Rio de Janeiro, Cabo Frio and Campos de Goytacazes counties.}
    \label{fig:map}
\end{figure}

\begin{figure}[ht!]
    \centering
    \includegraphics[width=0.85\textwidth]{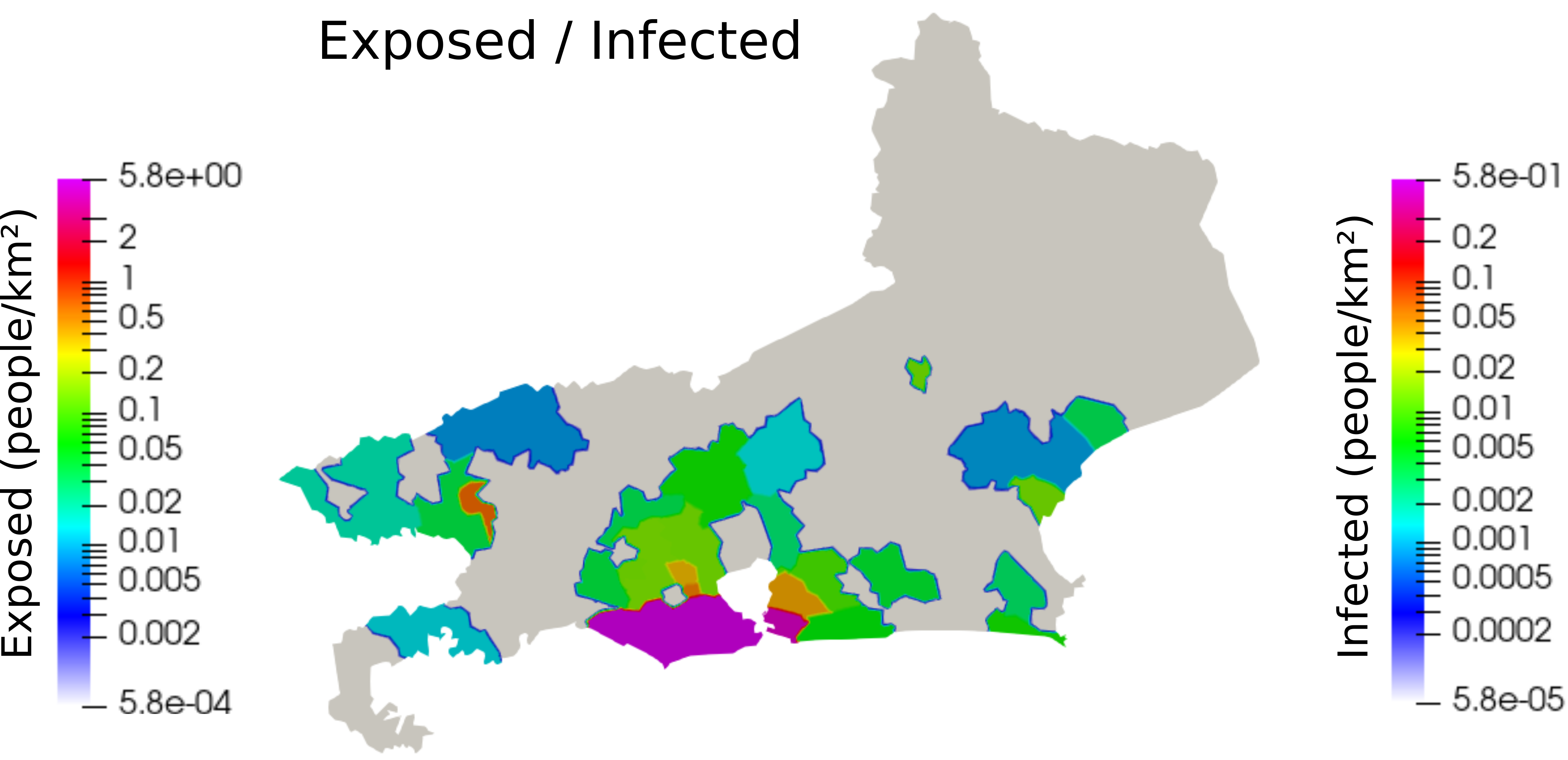}
    \caption{Initial exposed/infected population at Rio de Janeiro state.}
    \label{fig:initialexpinf}
\end{figure}

\begin{figure}[ht!]
    \centering
    \includegraphics[width=0.7\textwidth]{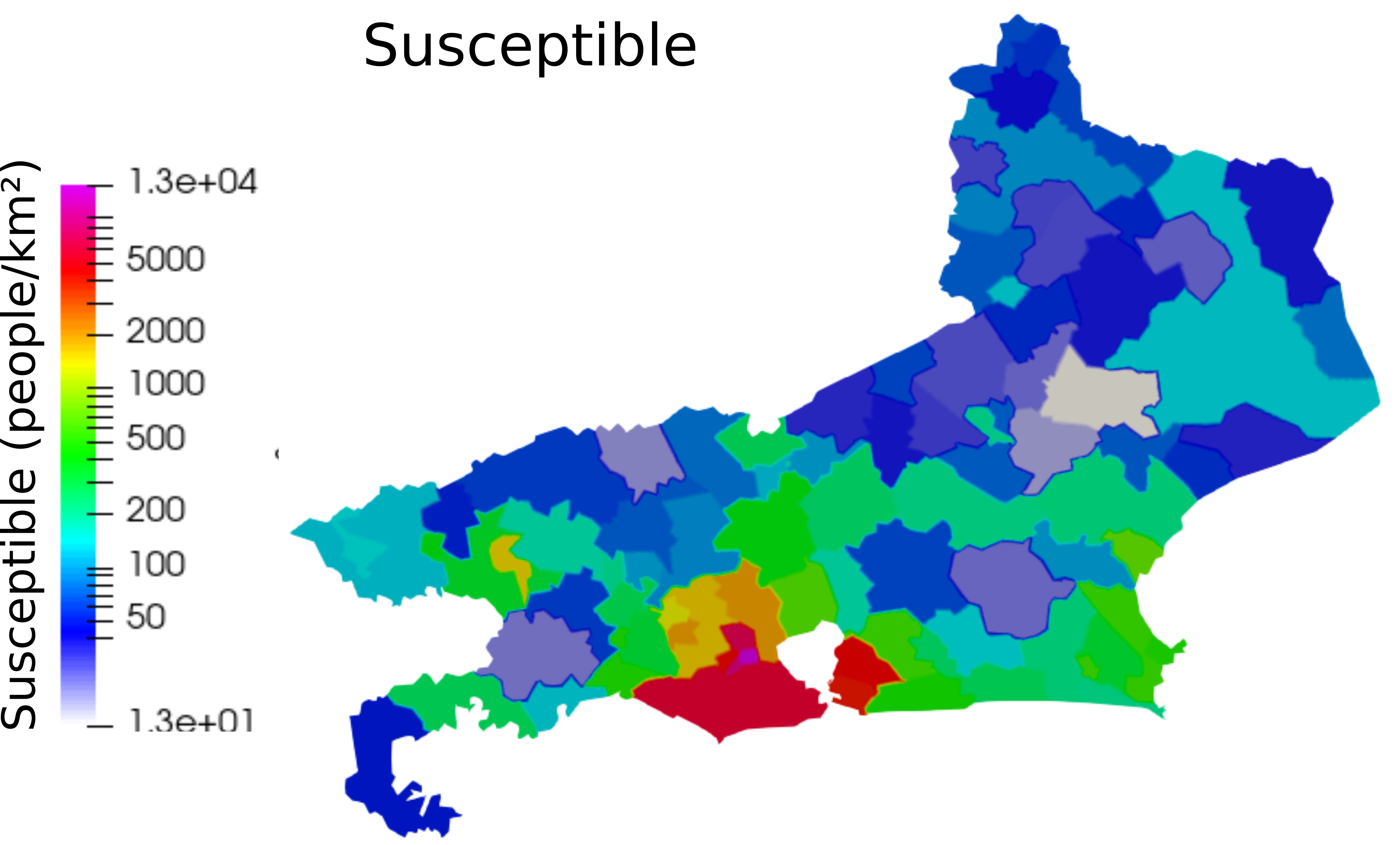}
    \caption{Initial susceptible population at Rio de Janeiro state.}
    \label{fig:initialsus}
\end{figure}

In \cite{grave2021assessing} we have simulated the spread of COVID-19 in the state of Rio de Janeiro, Brazil. We have shown difficulties in reproducing the virus spread dynamics in some counties, particularly in Cabo Frio (Region 2) and Campos dos Goytacases (Region 3), for example (see Fig. \ref{fig:map} for their locations). In that simulation, there were no initial infected or exposed populations in these counties (Fig. \ref{fig:initialexpinf}), making diffusion the only possible way for the virus to reach those areas. Given the large distance between them and the hotspots near the city center of Rio de Janeiro (Region 1, respectively 157/279 km distance from Regions 2 and 3), as well as the large sparsely-populated areas between them (see Fig. \ref{fig:initialsus}), population-weighted diffusion is not able to properly represent these dynamics. Note that, for all simulations considered in the following, parameters unrelated to the population transfer network are exactly as considered in \cite{grave2021assessing}. 

Cabo Frio (Region 2) is a coastal county, where there are seasonal visitors during the weekends. Therefore, we define $\eta_{1}^{2}$ as  people going from Rio de Janeiro to Cabo Frio on Fridays and coming back on Sunday, as in the previous simple test problem. The main difference is the percentages of people that comes and goes from each city. Since Rio has 6.7 million inhabitants and Cabo Frio has about 200 thousand, movement of the same amount of persons will correspond to different percentages of the population. We assume that about 30,000 people have this weekly pattern movement, or approximately 15\% of the population of Cabo Frio and 0.5\% of the population of Rio de Janeiro.
\begin{equation}\label{CaboToRio1}
    \eta_{1}^{2} = \begin{cases} 
    0.005/\text{day} \text{ if Friday,} \\
    0/\text{day} \text{ otherwise}
        \end{cases}, \qquad     \eta_{2}^{1} = \begin{cases} 
    0.15/\text{day} \text{ if Sunday,} \\
    0/\text{day} \text{ otherwise.}
        \end{cases}
\end{equation}
We apply the same rate to both $s$ and $e$ compartments. We note that, due to the previously-discussed population drifting effects, diffusion among the population, as well as the fact that these $\eta_i^j$ are applied to both $s$ and $e$, they may not correspond to exactly the same number of persons moved at each time period. Given that $s$ is several orders of magnitude larger than other compartments, the simplified rates defined above do not create serious issues in general. Indeed, these differences in number and composition of the population transferred in time may in fact be more realistic, as we would not expect such numbers to remain constant. Nevertheless, one may also define the $\eta_i^j$ similarly to \eqref{preciseEta} in order to ensure that population movements remain consistent in time.




Another region that was underestimated in the previous simulations for similar reasons to Cabo Frio is Campos dos Goytacazes (Region 3). We may then postulate movement between Rio de Janeiro and Campos dos Goytacazes, but consider a different weekly pattern, as Campos dos Goytacazes is not a vacation area. Rather, many people in the oil industry work there for one week in the offshore oil platforms, returning to Rio during the next week.  We assume that there are about 50,000 people who move in this manner, corresponding to 0.75\% the population of Rio de Janeiro, and 10\% the population of Campos dos Goytacazes, and adopt the convention of denoting \textit{even-numbered} and \textit{odd-numbered} weeks in the obvious way. Then, analogously to \eqref{CaboToRio1}:

\begin{equation}\label{RioToCampos1}
    \eta_{1}^{3} = \begin{cases} 
    0.0075 \cdot \text{week}^{-1} \text{ on odd-numbered weeks,} \\
    0 \cdot \text{week}^{-1} \text{ otherwise}
        \end{cases}, \qquad     \eta_{3}^{1} = \begin{cases} 
    0.1 \cdot \text{week}^{-1} \text{ on even-numbered weeks,} \\
    0 \cdot \text{week}^{-1} \text{ otherwise,}
        \end{cases}
\end{equation}
for both $\eta_{i}^{j,s}$ and $\eta_{i}^{j,e}$. 



Lastly, to complete the network, we also consider movement between Campos dos Goytacases and Cabo Frio. We define this movement analogously to \eqref{CaboToRio1}; however, rather than moving approximately 30,000 people between the regions, we move about 10,000, corresponding to about 2\% of the population of Campos dos Goytacases and 4.45\% of the population of Cabo Frio:

\begin{equation}\label{CaboToCampos1}
    \eta_{3}^{2} = \begin{cases} 
    0.02/\text{day} \text{ if Friday,} \\
    0/\text{day} \text{ otherwise}
        \end{cases}, \qquad     \eta_{2}^{3} = \begin{cases} 
    0.0445/\text{day} \text{ if Sunday,} \\
    0/\text{day} \text{ otherwise.}
        \end{cases}
\end{equation}



\begin{figure}[ht!]
    \centering
    \includegraphics[width=\textwidth]{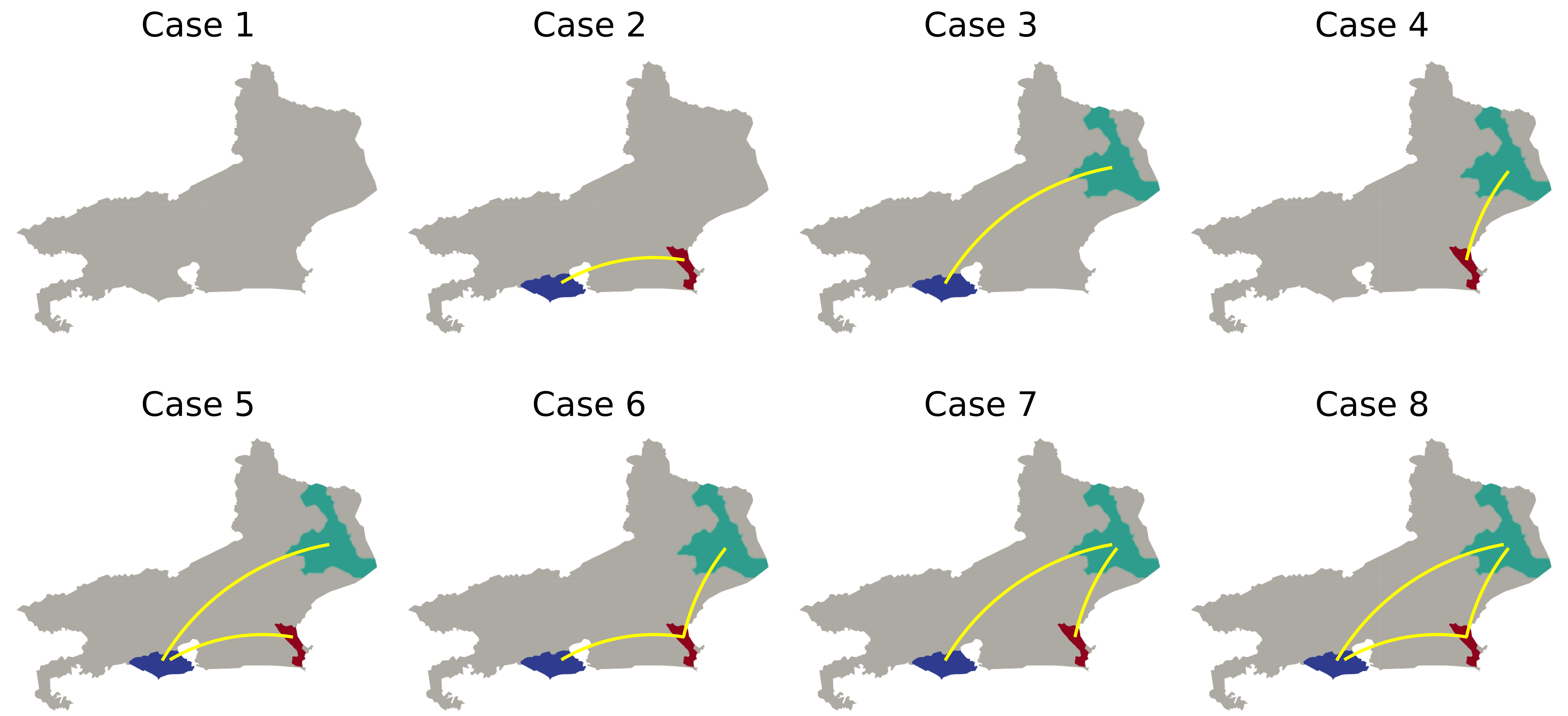}
    \caption{Definition of the movement for each case. The region in blue is Rio de Janeiro (RJ), red is Cabo Frio (CF) and green is Campos dos Goytacazes (CG).}
    \label{fig:cases}
\end{figure}

In total, we simulate eight different cases, considering different networks as presented in Table \ref{tab:cases} and Figure \ref{fig:cases}. The mesh of the Rio de Janeiro region contains  11,632 elements and 6,082 nodes, and, although unstructured, the element size is very close to uniform, with a mean area of 1 km$^2$ and 95\% of elements between 0.5 and 1.5 km$^2$. Simulations are done considering linear elements.

\begin{table}[ht!]
    \centering
\begin{tabular}{ |c|c|c|c|c| } 
\hline
& & \multicolumn{3}{|c|}{Movement between} \\
 Case & PDE &  RJ and CF & RJ and CG & CF and CG\\ 
 \hline
 1 & X & & & \\ 
  \hline
 2 & X & X & & \\ 
  \hline
 3 & X &  & X &  \\ 
  \hline
 4 & X &  & & X \\ 
  \hline
 5 & X & X & X &  \\ 
  \hline
 6 & X & X & & X \\ 
  \hline
 7 & X &  & X & X \\ 
  \hline
 8 & X & X & X & X \\ 
 \hline
\end{tabular}
    \caption{Definition of the movement between Rio de Janeiro (RJ), Cabo Frio (CF) and Campos dos Goytacazes (CG) for each case.}
    \label{tab:cases}
\end{table}

%

\begin{figure}[ht!]
    \centering
    \includegraphics[width=0.7\textwidth]{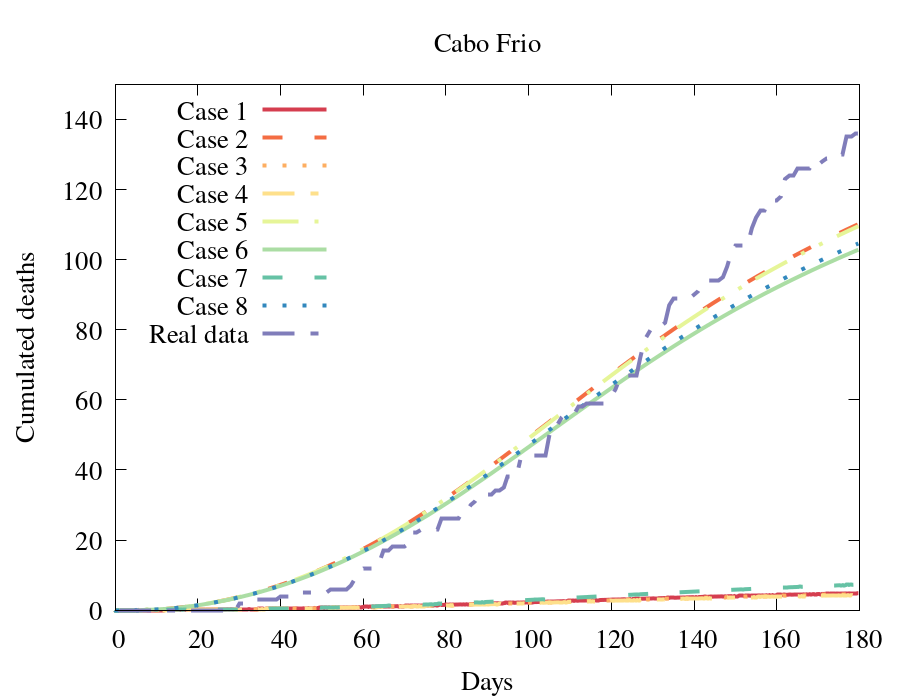}
    \caption{Cabo Frio cumulative deaths}
    \label{fig:cabofrio}
\end{figure}

\begin{figure}[ht!]
    \centering
    \includegraphics[width=0.7\textwidth]{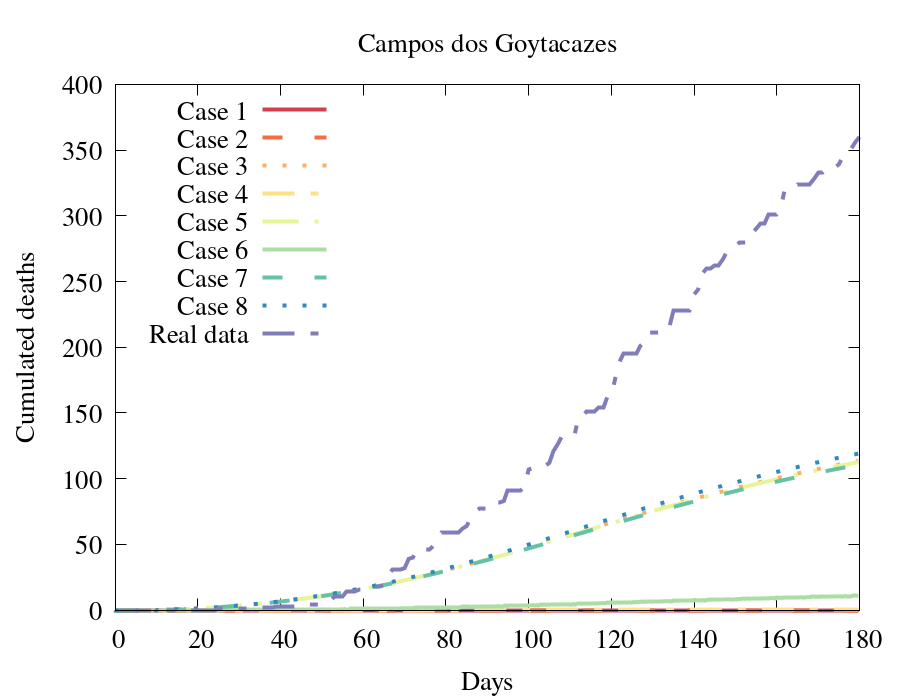}
    \caption{Campos dos Goytacazes cumulative deaths}
    \label{fig:campos}
\end{figure}

\begin{figure}[ht!]
    \centering
    \includegraphics[width=0.7\textwidth]{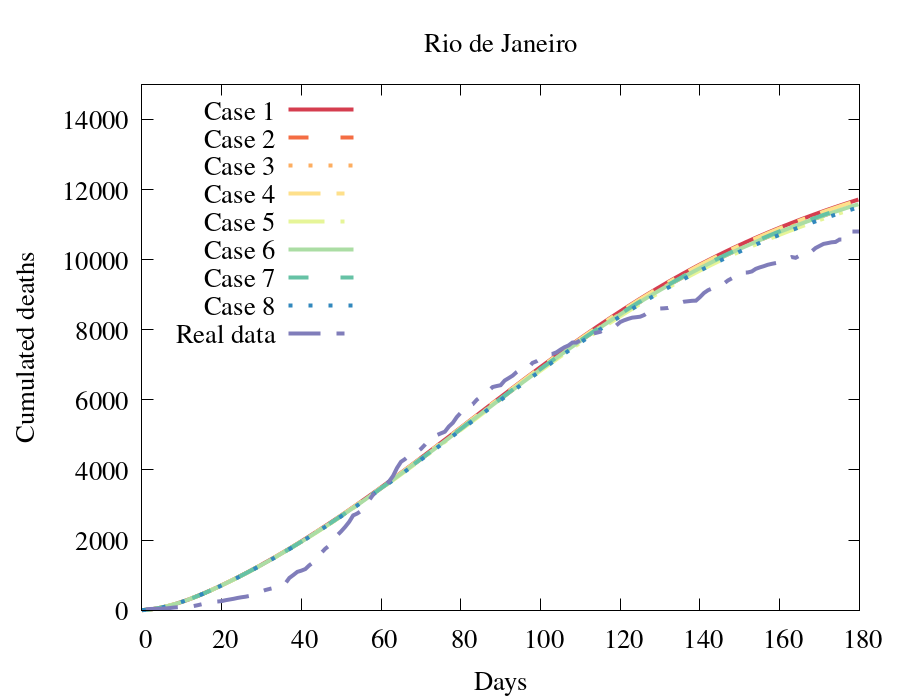}
    \caption{Rio de Janeiro cumulative deaths}
    \label{fig:rio}
\end{figure}

\begin{figure}[ht!]
    \centering
    \includegraphics[width=\textwidth]{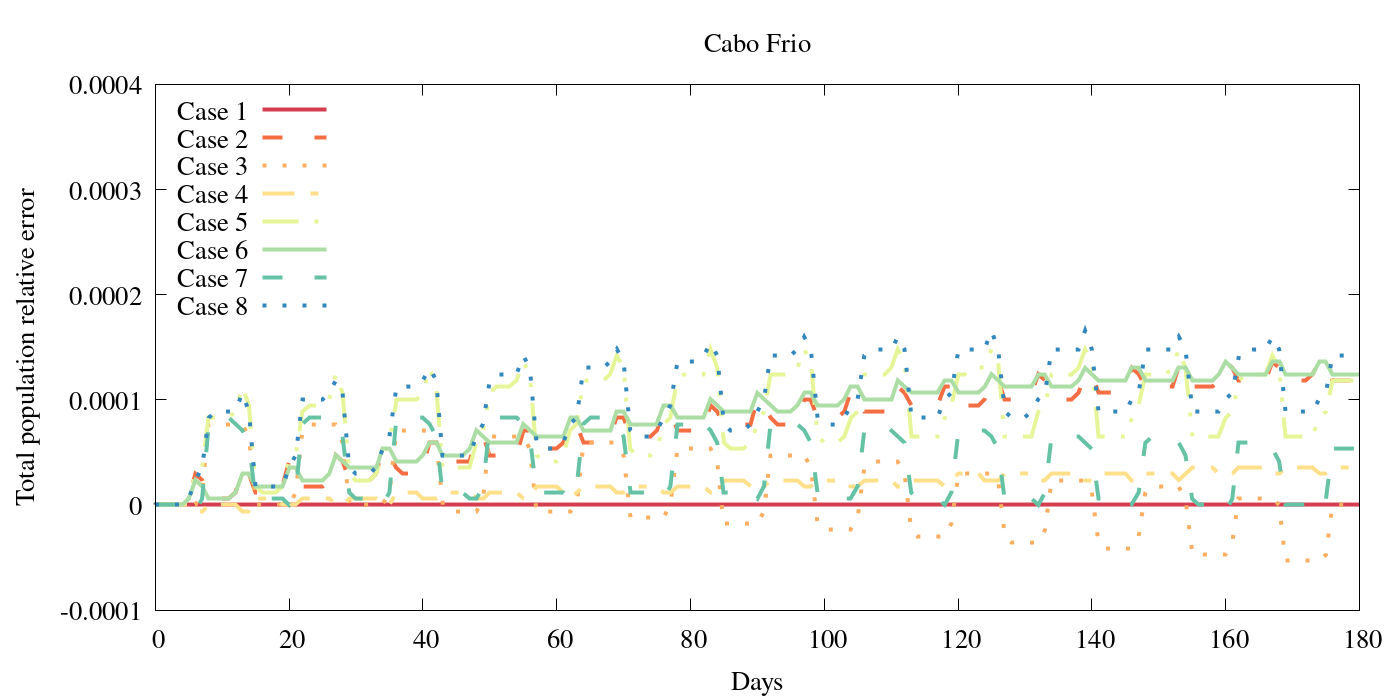}
    \caption{Total population relative error considering population movement}
    \label{fig:masscabofrio}
\end{figure}

In Figures \ref{fig:cabofrio}, \ref{fig:campos} and \ref{fig:rio} we show the comparison of cumulative deaths between the observed data and the simulated cases. For each case, we are considering diffusion. Hence, we do not have an analytic solution as in the previous example. However, we do know that the total overall population in the region should be conserved. In Figure \ref{fig:masscabofrio} we show the total population relative error with time, which is substantially under 1\% throughout the time period. 

When considering isolated movement between Cabo Frio and Campos dos Goytacazes (Case 3) we see very little contribution to the spread of the virus in these counties because they both have no infected or exposed at the beginning of the simulation, and no nonlocal propagation of the virus occurs. However, when incorporating movement from Rio de Janeiro, which has many exposed and infected in the early stages of the epidemic (as in Cases 1 and 2), we see a marked improvement in the simulated agreement with reality. In particular, we observe that the nonlocal movement defined by the population transfer network enables the epidemic to spread in these regions. We also note that, as the population in Rio de Janeiro is much larger than in Cabo Frio or Campos dos Goytacazes, the deceased population in Rio shows little change when introducing nonlocal population movement.

\begin{figure}[ht!]
    \centering
    \includegraphics[width=\textwidth]{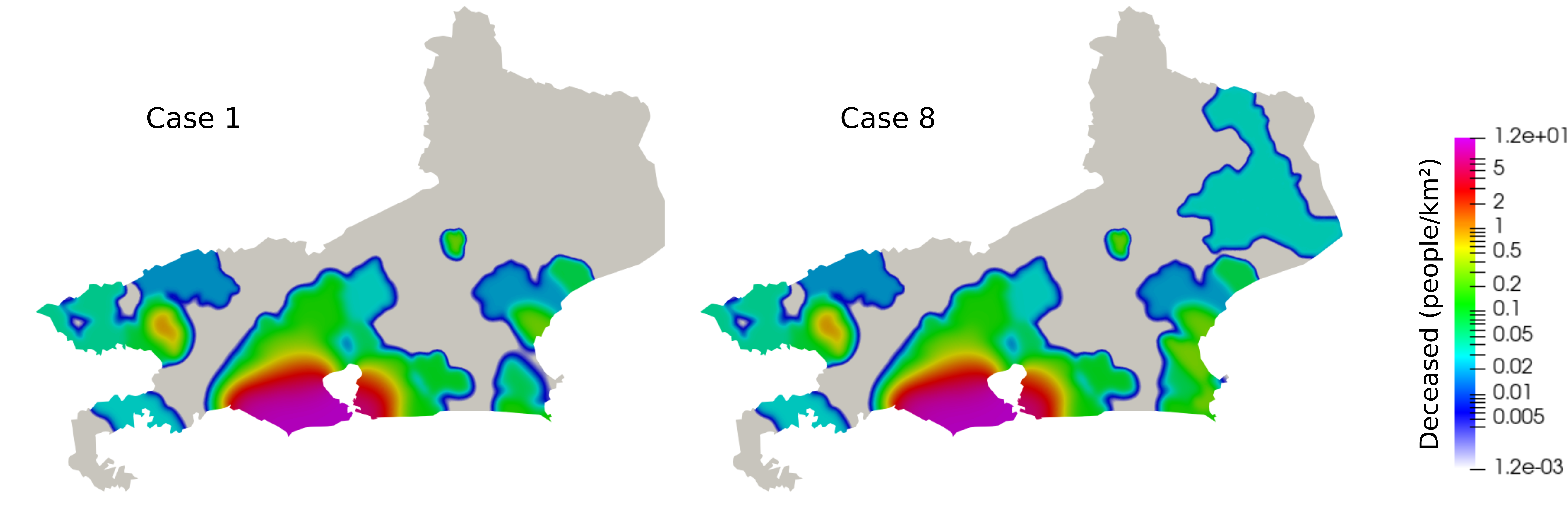}
    \caption{Spatial distribution of the deceased population at t=180 days for Case 1 (no population movement) and Case 8 (population movement network between RJ, CF and CG).}
    \label{fig:deceased}
\end{figure}

To compare the effects of the population transfer network, Figure \ref{fig:deceased} shows the spatial distribution of the deceased population at t=180 days for Case 1 (no population transfer) and Case 8 (full network incorporating population transfer between Rio de Janeiro, Cabo Frio, and Campos dos Goytacazes). In Rio de Janeiro, we see that the purely diffusive model (Case 1) already works well, and the introduction of the population transfer network does not cause any sort of deterioration in accuracy (Case 3). In contrast, in Cabo Frio and Campos dos Goytacazes, the purely diffusive model fails to recreate the observed dynamics. However, upon the introduction of the population transfer network, in Cabo Frio, the results closely match the observed data. In the case of Campos dos Goytacazes, the simulations now match measured data more closely, as the deaths are no longer close to zero; however, they are still a fair deal lower than the observed reality. This may be due to our currently simplistic assumptions about the nature of the population movement; under more realistic assumptions, these results may improve.
\section{Discussion and Future Directions}
In the present work, we have introduced a new approach to modeling nonlocal dynamics in reaction-diffusion models of infectious disease transmission. The proposed method involves defining different subregions in the domain of interest as nodes in a network, with the movement between nodes defined via edges in a time-varying graph. While conceptually simple, defining an operator properly representing these dynamics within a reaction-diffusion PDE is nontrivial. We discussed the proper definition of such an operator that both respects sensible physical constraints and maintains the mathematical well-posedness of the underlying system. In particular, the operator must be non-negative and population (mass)-conservative. We then demonstrated conditions that ensure that the transfer operator satisfies these important properties, and proposed some natural candidates for its definition.
\par We then performed two numerical experiments with different goals. The first was designed to confirm that our numerical implementation of the network transfer operator demonstrated the necessary conservation and non-negativity properties as outlined in Section 3. We found that our implementation generally respects conservation, with numerical errors below one percent for all meshes considered, and tending to zero with mesh refinement. Unsurprisingly, we also found that structured meshes generally demonstrate superior conservation properties.
\par Our second numerical experiment was a proof-of-concept designed to demonstrate the potential of this approach on a realistic problem. We considered a three-node network of three distinct regions in the state of Rio de Janeiro in Brazil. Without considering population transfer along the network, the PDE model is able to accurately capture the dynamics in the Rio de Janeiro region. However, it is unable to account for the contagion in the Cabo Frio and Campos dos Goytacazes regions, due to the inherent lack of nonlocal dynamics. We then showed that, under reasonable assumptions, introducing a network structure endowed with a population transfer operator is able to account for these nonlocal dynamics, improving the model's performance in the Campos de Goytacazes and Cabo Frio regions.
\par There are several important directions for future research. We considered only the uniform distribution of the population transfer operator in the present work; however, it is important to know the sensitivity of the model to this specific definition. Given the observed mesh dependence on population conservation, it may also be beneficial to explore adaptive meshing algorithms that take the geography of the network regions into account, as this may improve performance. Perhaps most importantly, we stress that the simulations shown here were also preliminary and designed primarily to show the potential of this methodology to account for nonlocal dynamics, and the considered network was very simple. However, applying the described approach to more complex networks with more nodes is important. Similarly, we defined the edges in a rudimentary way. In future work, these edges must be defined more realistically, incorporating measured data. Such extensions represent an important step toward the development of a realistic model which may ultimately inform public health decision-making.
\label{sec:discussion}

\section*{Acknowledgments}
The work was partially supported by the Brazilian agencies CAPES (Finance code 001), FAPERJ (ALGAC), and CNPQ (ALGAC and RFSA). M. Grave was partially supported by Fiocruz. A. Reali was partially supported  by the Italian Ministry of University and Research (MIUR) through the PRIN project XFAST-SIMS (No. 20173C478N).


\bibliographystyle{abbrv}  
\bibliography{references}

\end{document}